\documentclass[twocolumn]{aastex62}
\begin{document}
\title{Precursor Wave Amplification by Ion-Electron Coupling through
Wakefield in Relativistic Shocks}

\author{Masanori Iwamoto}
\affiliation{Department of Earth and Planetary Science, University of Tokyo,
7-3-1 Hongo, Bunkyo-ku, Tokyo 113-0033, Japan}

\author{Takanobu Amano}
\affiliation{Department of Earth and Planetary Science, University of Tokyo,
7-3-1 Hongo, Bunkyo-ku, Tokyo 113-0033, Japan}

\author{Masahiro Hoshino}
\affiliation{Department of Earth and Planetary Science, University of Tokyo,
7-3-1 Hongo, Bunkyo-ku, Tokyo 113-0033, Japan}

\author{Yosuke Matsumoto}
\affiliation{Department of Physics, Chiba University, 1-33 Yayoi, Inage-ku,
Chiba, Chiba 263-8522, Japan}

\author{Jacek Niemiec}
\affiliation{Instytut Fizyki J\c{a}drowej PAN, ul. Radzikowskiego 152, 31-342
Krak\'ow, Poland}

\author{Arianna Ligorini}
\affiliation{Instytut Fizyki J\c{a}drowej PAN, ul. Radzikowskiego 152, 31-342
Krak\'ow, Poland}

\author{Oleh Kobzar}
\affiliation{Instytut Fizyki J\c{a}drowej PAN, ul. Radzikowskiego 152, 31-342
Krak\'ow, Poland}

\author{Martin Pohl}
\affiliation{Institut f\"ur Physik und Astronomie, Universit\"at Potsdam
Karl-Liebknecht-Strasse 24/25, 14476 Potsdam, Germany}
\affiliation{DESY, Platanenallee 6, 15738 Zeuthen, Germany }

\correspondingauthor{Masanori Iwamoto}
\email{iwamoto@eps.s.u-tokyo.ac.jp}

\begin{abstract}

 We investigated electromagnetic precursor wave emission in relativistic
 shocks by using two-dimensional particle-in-cell simulations.
 We found that the wave amplitude is significantly enhanced by a positive
 feedback process associated with ion-electron coupling through the
 wakefields for high magnetization.
 The wakefields collapse during the nonlinear process of the parametric
 decay instability in
 the near-upstream region, where nonthermal electrons and ions are
 generated. The intense coherent emission and the particle acceleration
 may opperate in high-energy astrophysical objects.

\end{abstract}

 \keywords{cosmic rays --- High energy astrophysics --- plasma physics --- shocks}

\section{Introduction}\label{sec:intro}

The acceleration mechanism for generating ultra-high-energy cosmic rays
(UHECRs) with energies above $10^{18}$ eV is one of the most important
unsolved problems in astrophysics. Relativistic shocks in extragalactic
astrophysical objects such as jets from active galactic nuclei and
gamma-ray bursts are considered as efficient acceleration sites
\citep[e.g.,][]{Piran2005,Marscher2006} and are thus likely sources for
UHECRs
\citep[e.g.,][]{Hillas1984,Biermann1987,Milgrom1995,Vietri1995,Waxman1995}.
Observations of anisotropy in the arrival direction of the UHECRs favor
the extragalactic origin \citep[e.g.,][]{Abbasi2014,Aab2015,Aab2017,Aab2018}.
In addition, the recent observation by IceCube reported that a high-energy
neutrino event associated with the UHECRs correlates in direction and
time with a gamma-ray flare from the blazar
\citep{Aartsen2018}.

In relativistic perpendicular shocks, large-amplitude electromagnetic waves
are excited by synchrotron maser instability (SMI).
Since the precursor waves are intense and coherent,
fast radio bursts (FRBs) may be attributed to the SMI in relativistic shocks
\citep[e.g.,][]{Lyubarsky2014,Plotnikov2019,Metzger2019}.
\added{The SMI induces extraordinary mode (X-mode) waves and thus
the polarization of the precursor waves also supoorts the FRB model
\cite[see][]{Plotnikov2019}.}
The wave emission has been widely studied by means of
one-dimensional (1D) particle-in-cell (PIC) simulations
\citep[e.g.,][]{Langdon1988,Hoshino1991,Gallant1992,Hoshino1992,Amato2006}.
\cite{Lyubarsky2006} showed that electrons lag behind ions
inside such large-amplitude wave and that a longitudinal electric field
is excited in the wake of the precursor wave. \cite{Hoshino2008}
extended the work and demonstrated that the pump electromagnetic wave
decays into a Langmuir wave via parametric decay instability (PDI)
\citep[e.g.,][]{Mima1984,Kruer1988}.
\added{PDI is a wave-wave interaction and large-amplitude
electromagnetic waves are subject to it because they exert radiation
pressure on plasmas and induce forward-propagating compressive fluctuations.
When Raman scattering which is an inelastic scattering process of light
works in plasmas, the PDI produces Langmuir waves.}
He found that nonthermal
particles are generated in the manner analogous to wakefield
acceleration (WFA) during the nonlinear process of the Langmuir wave
collapse. The WFA is first proposed in laboratory plasmas
\citep{Tajima1979} and later applied to UHECR acceleration
\citep[e.g.,][]{Chen2002, Arons2003, Murase2009}.
\added{In laboratory plasmas, wakefield, which is a londitudinal
electrostatic wave, is excited by an ultra-intense laser pulse. Some
electrons are preferentially accelerated by the wakefield via Landau
resonance.}
Previous laser-plasma experiments \citep{Kuramitsu2011a,Kuramitsu2011b} and
simulations \citep{Kuramitsu2008,Kuramitsu2012,Liu2017,Liu2018,Liu2019}
showed that the WFA produces a power-law distribution with a spectral index
of 2. Therefore, the WFA in the context of relativistic shocks is a
promising candidate for UHECR acceleration.

Recently, our high-resolution two-dimensional
(2D) PIC simulations in pair plasmas \citep{Iwamoto2017,Iwamoto2018}
showed that the wave emission continues even well after Weibel instability
\citep{Weibel1959,Fried1959} becomes active at the shock front. The precursor
waves are large enough to disturb the upstream plasma and induces transverse
density filaments extended well ahead of the shock front
\citep[see also][]{Plotnikov2018}. The interaction between the strong
precursor
waves and the upstream plasma is not negligible even in pair plasmas
\citep[see also][]{Lyubarsky2018}. Although our results for pair plasmas are
favorable for the WFA scenario in relativistic shocks, we
could not directly demonstrate it because finite mass difference of two
opposite charges is essential for exciting the wakefield.

In this letter, by performing 2D PIC simulation of ion-electron shocks,
we demonstrate that the wakefield is indeed induced by the large-amplitude
precursor waves. Especially for high magnetization, the wave amplitude is
significantly amplified due to a positive feedback process associated with
ion-electron coupling. Nonthermal electrons and ions are generated
during the nonlinear collapse of the wakefield and the particle
energy spectrum shows a power-law distribution.
Our self-consistent simulations of multidimensional relativistic shocks
provide insights into the physics of wave-plasma interaction and particle
acceleration in high-energy astrophysical objects.

\section{simulation Setup}\label{sec:setup}

We employed a fully relativistic electromagnetic PIC code
\citep{Matsumoto2013,Matsumoto2015}, which enables us to follow long-term
evolution by minimizing the effect of numerical Cherenkov instability
\citep{Ikeya2015}. The basic configuration is nearly identical to our
previous simulations \citep{Iwamoto2017}. The essential difference is a
finite ion-to-electron mass ratio $m_i/m_e = 50$. The simulation domain is
in the $x$--$y$ plane and the periodic boundary condition is applied in the
$y$ direction. We consider purely perpendicular shocks and the ambient
magnetic field $B_1$ is in the $z$ direction. The cold
plasma flow (with zero thermal spread of electrons and ions) is
injected from the right-hand boundary with the bulk Lorentz
factor $\gamma_1=40$ toward the left-hand reflecting wall at
$x=0$. The interaction between incoming and reflected particles triggers
the shock propagating toward $+x$ direction. The upstream particle
number per cell $N_1$ is set as $N_1\Delta x^2 = 64$ for each particle
species. The gird size $\Delta x$ and the time steps $\Delta t$ are
fixed to $\Delta x/(c/\omega_{pe})=1/40$ and $\omega_{pe}\Delta t=1/40$,
where $\omega_{pe}=\sqrt{4\pi N_1e^2/\gamma_1m_e}$ is the proper electron
plasma frequency.
Note that our simulations with the CFL number
$c\Delta t/\Delta x = 1$ are numerically stable because of an implicit
Maxwell solver used in our code.
The basic structure of relativistic magnetized shocks is
characterized by the ratio of the Poynting flux to the upstream bulk
kinetic energy flux $\sigma_s = B_1^2/4\pi\gamma_1N_1m_sc^2$,
where the subscript $s=i,e$ represents particle species. Our shock
simulations are performed for values of $\sigma_e = 5$, $1$, and $0.1$.
Note that $\sigma_i = (m_e/m_i) \sigma_e$ is always satisfied
and $\sigma_i$ is then $0.1$, $2 \times 10^{-2}$, and
$2 \times 10^{-3}$, respectively.
The grid size in the unit of the electron gyroradius is given by
$\Delta x/(c/\omega_{ce})=\sqrt{\sigma_e}/40$, where
$\omega_{ce}=eB_1/\gamma_1m_ec = \sqrt{\sigma_e}\omega_{pe}$ is the
electron cyclotron frequency, and thus it varies from $\sim 0.06$ at
$\sigma_e = 5$ to $\sim 0.008$ at $\sigma_e = 0.1$.
We used the different simulation box sizes for each magnetization
and the number of grids is $N_x \times N_y = 80000 \times 1600$, $80000
\times 800$, and $200000 \times 1600$, respectively.

\added{The ambient magnetic fied is fixed in the $z$ direction
(i.e., out-of-plane configuration) thorughout this study.
\cite{Iwamoto2018} show that in the in-plane configuration,
ordinary mode (O-mode) waves as well as X-mode waves
are excited due to fluctuations along the ambient magnetic field. For
$\sigma_e \gtrsim 10^{-1}$, the X-mode waves are dominant
over the O-mode waves and the X-mode wave amplitude is almost equal to that
in the out-of-plane configuration. Therefore, we focus on the out-of-plane
configuration in this study.}

\section{Shock Structure}\label{sec:structure}

\begin{figure*}[htb!]
 \plotone{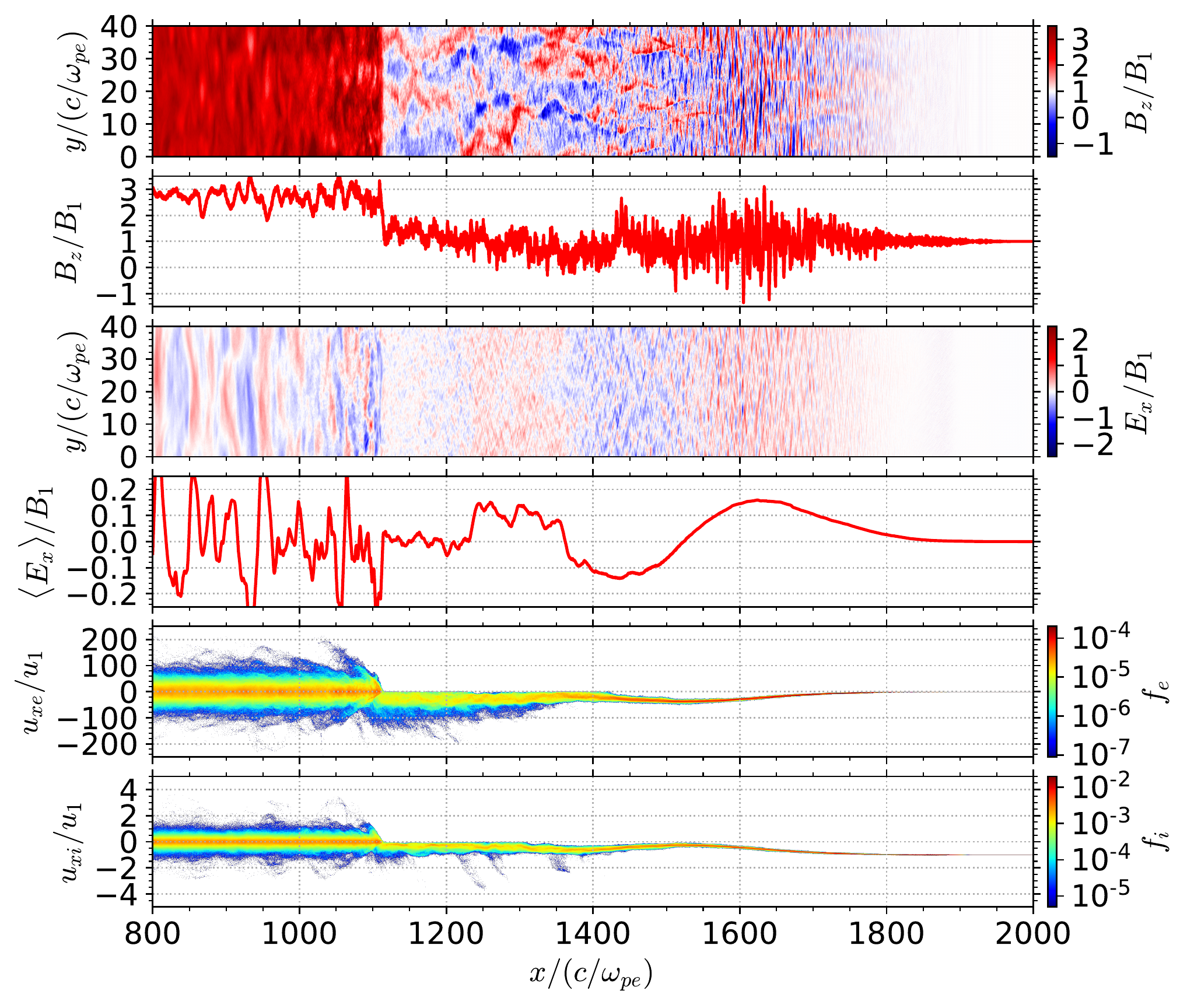}
 \caption{Shock structure for $\sigma_e=5$ at $\omega_{pe}t=2000$. The
 out-of-plane magnetic field $B_z$, the 1D profile of $B_z$, the longitudinal
 electric field $E_x$, the $y$-averaged electric field $\langle E_x \rangle$,
 and the phase space plots in the $x$--$u_{xs}$ plane for electrons and ions
 are shown.}
 \label{highsig}
\end{figure*}

Figure \ref{highsig} shows the global structure of the shock
at $\omega_{pe}t=2000$ in the case of relatively high magnetization:
$\sigma_e = 5$. From top to bottom, the
out-of-plane magnetic field $B_z$, the 1D cut of $B_z$ at
$y/(c/\omega_{pe}) = 20$, the longitudinal electric field $E_x$, the
$y$-averaged electric field $\langle E_x \rangle$, and the electron and
ion phase space densities $x-u_{xs}$ integrated over the $y$ direction are
shown. Here $u_1 = \sqrt{\gamma_1^2-1} \simeq 40$ is the bulk four velocity
of injected particles. The large-amplitude electromagnetic waves are
clearly seen in the upstream. The precursor wave induces a large-scale
longitudinal electric field, which is so-called wakefield as already
discussed by \cite{Hoshino2008}. The linear theory of the PDI via
forward Raman scattering process shows that the wavelength of the Langmuir
wave measured in the simulation frame can be estimated by
$\lambda_L/(c/\omega_{pe})\simeq 4\pi \gamma_1 \simeq 500$
\citep[see, e.g.,][]{Mima1984,Kruer1988,Hoshino2008}.
This estimate comes from the frequency and wavevector matching
conditions of the three waves under the assumption that the pump wave
frequency is sufficently large.
Considering the sinusoidal part of the wakefield in the region
$1350 \lesssim x/(c/\omega_{pe}) \lesssim 1850$,
this estimate gives a good agreement with our simulation result.
The wakefield breaks up in the region
$1150 \lesssim x/(c/\omega_{pe}) \lesssim 1350$, indicating that the PDI
enters the nonlinear phase. The precursor waves also become turbulent there
and the transverse filamentary structure is generated.
Both electrons and ions are strongly heated and accelerated in the turbulent
region.

\section{Precursor Wave amplification}\label{sec:amp}

\begin{figure}[htb!]
 \plotone{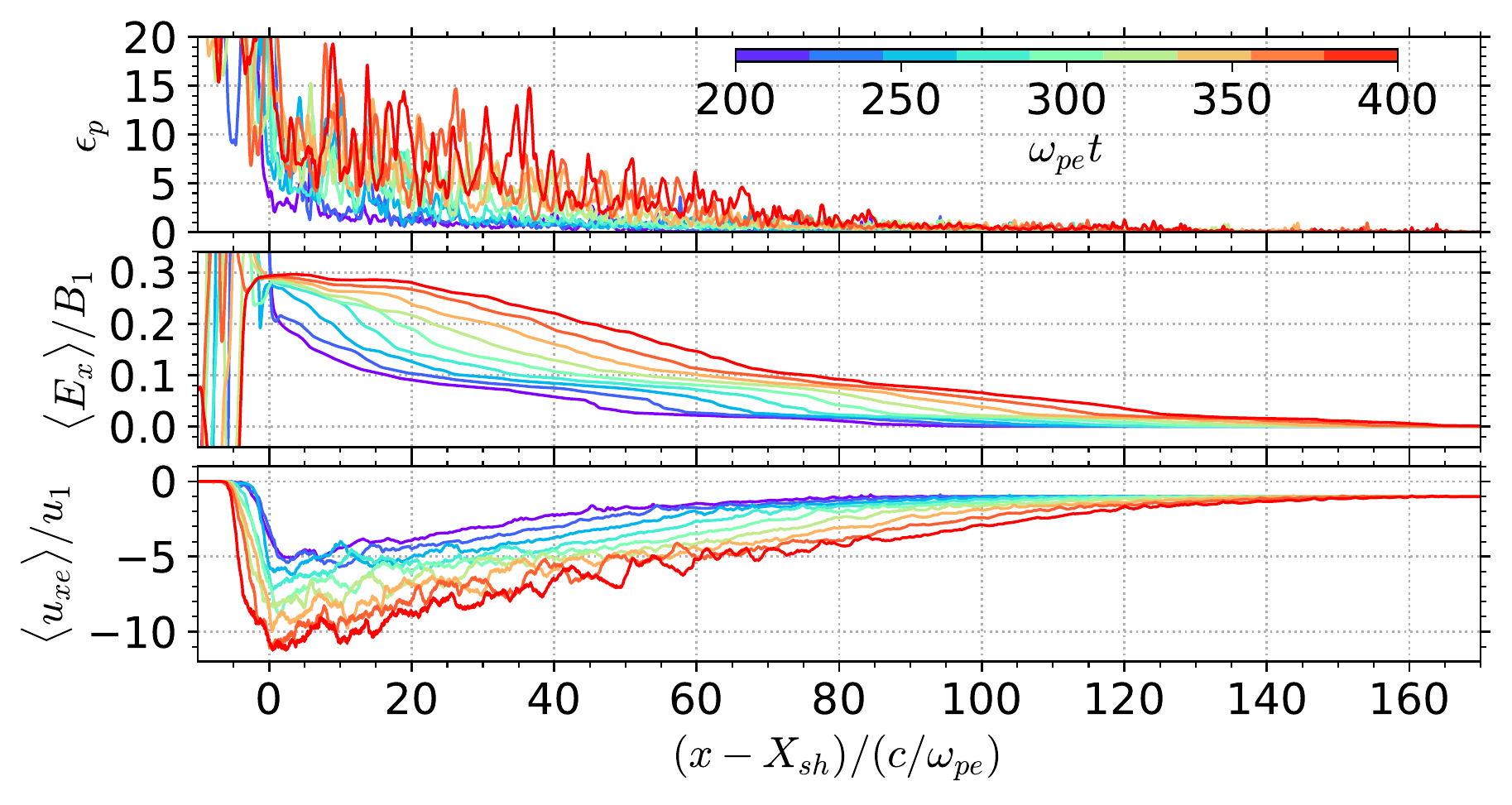}
 \caption{Temporal evolution of near-upstream shock structure. The
  precursor wave energy $\epsilon_p $, the wakefield
  $\langle E_x  \rangle$, and the electron bulk velocity
  $\langle u_{xe} \rangle$ are shown form top to bottom.}
 \label{evo}
\end{figure}

The modulation of the precursor wave amplitude is clearly visible in
Figure \ref{highsig}, indicating that the amplitude changes in time. Figure
\ref{evo} shows temporal evolution of the precursor wave energy normalized
by the upstream electron kinetic energy
$\epsilon_p = \delta B^2/4\pi \gamma_1N_1m_ec^2$, the
wakefield $\langle E_x \rangle$, and the electron bulk velocity
$\langle u_{xe} \rangle$.
Here $\delta B = B_z-B_1$ is the wave component of the magnetic field.
Note that postive longitudinal electric fields are excited in an
initial phase of the PDI because electrons lag behind ions inside
intense precursor waves \citep[see][]{Lyubarsky2006}.
One can find that $\epsilon_p$ increases in time and that the wave
amplification are coincident with increase of the magnitude of
$\langle E_x \rangle$ and $\langle u_{xe} \rangle$.
This observation indicates that the accelerated electrons cause the increase
of the wave amplitude which initiates further acceleration of electrons
through the wakefield, and that the positive feedback process discussed in
the earlier studies \citep{Lyubarsky2006,Hoshino2008} operates in 2D as well.
Note that the wave emission efficiency is mainly controlled by electrons
because ions do not contribute to the electromagnetic wave emission via the
SMI \citep{Hoshino1991}.
Initially, the precursor waves are emitted by consuming the free energy of
electrons with an initial Lorentz factor of $\gamma_1$. The precursor waves
then induce a wakefield via the PDI in the immediate upstream of the shock.
The incoming electrons are gradually accelerated or decelerated by the
wakefield on their way to the shock. When the accelerated electrons hit the
shock, the precursor wave may be amplified because of an increased free
energy available for the SMI.
The amplified precursor waves induce a stronger
wakefield which in turn accelerates the upstream incoming electrons
even further. If this self-reinforcing cycle operates sufficiently rapidly,
the precursor wave amplitude continuously grows over time.
The ion kinetic energy is converted into the electron through the wakefield
during this positive feedback process and the SMI indirectly consumes the
ion kinetic energy. The precursor wave amplitude takes
the maximum value when the electron kinetic energy achieves equipartition
with the ion \citep{Lyubarsky2006}.

\begin{figure*}[htb!]
 \plottwo{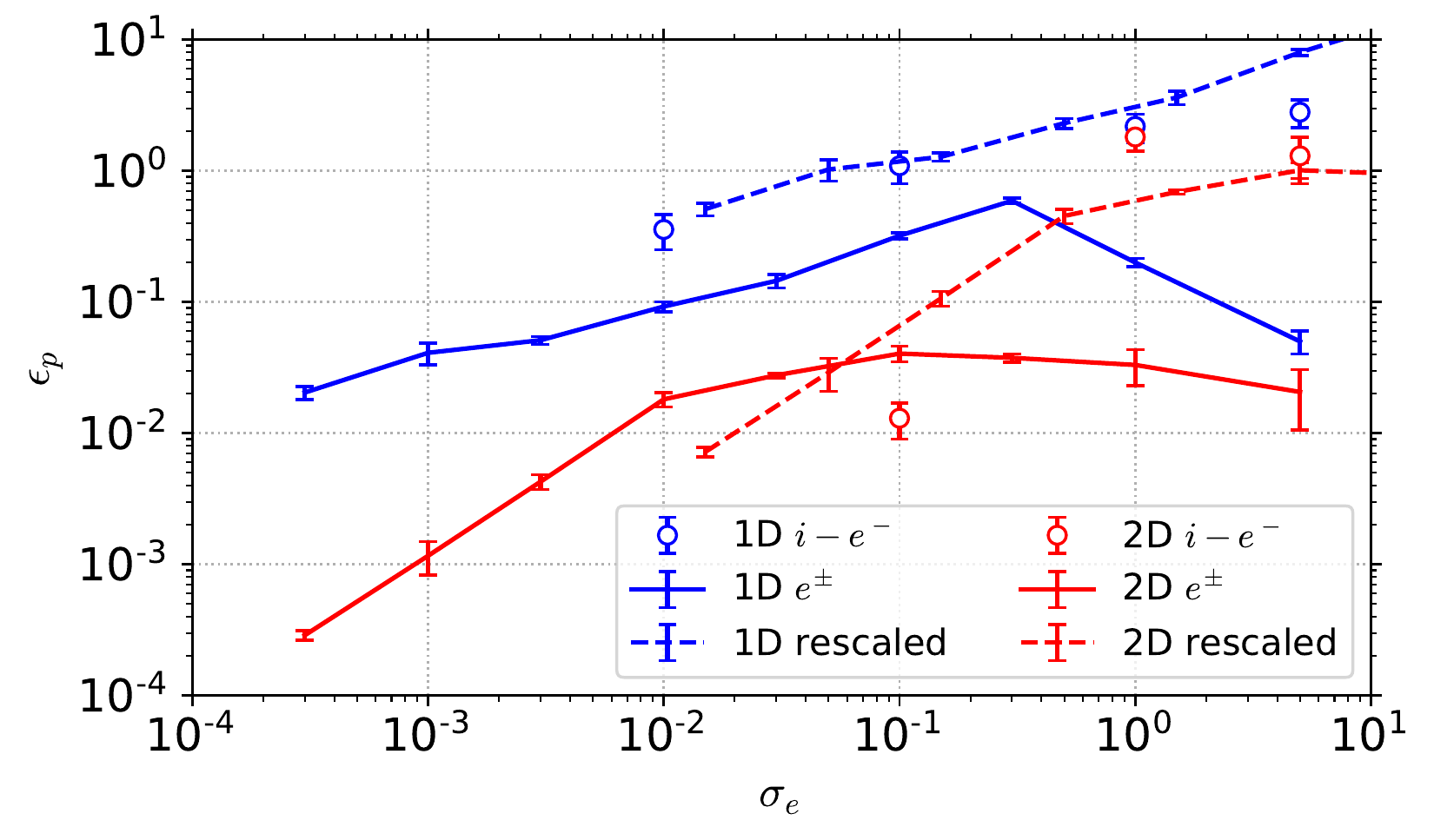}{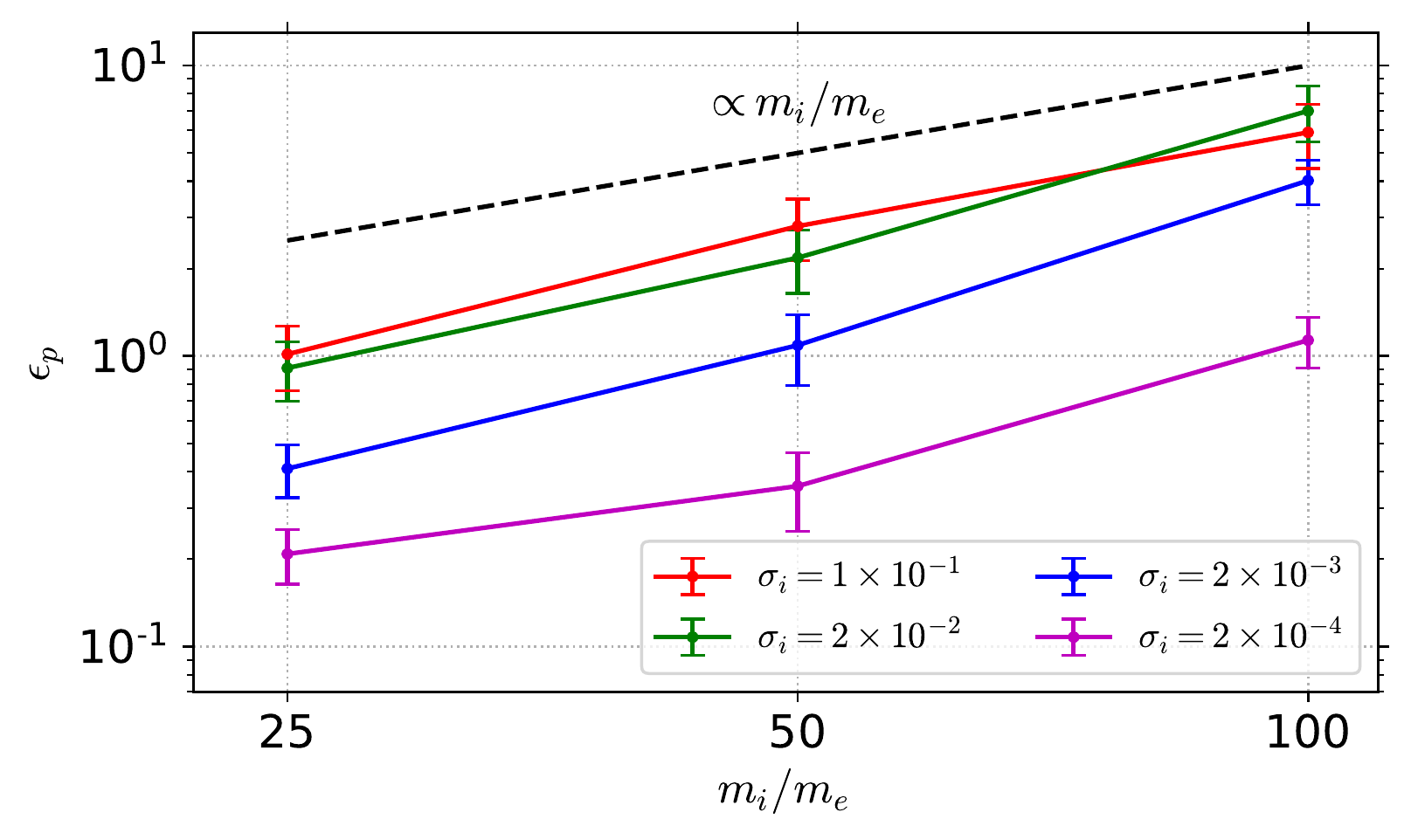}
 \caption{Left panel: The precursor wave energy $\epsilon_p$ in 1D (blue)
  and 2D (red) as a function of $\sigma_e$. The circles and solid lines
  indicate ion-electron and pair shock simulation results, respectively. The
  rescaled wave energy in pair plasmas is shown in the dashed lines.
  Right panel: The mass ratio dependence of the 1D wave energy
  $\epsilon_p$ at $\sigma_i = 1 \times 10^{-1}$ (red), $2 \times 10^{-2}$
  (green), $2 \times 10^{-3}$ (blue), and $2 \times 10^{-4}$ (magenta).}
 \label{amp}
\end{figure*}

Left panel of Figure \ref{amp} shows $\sigma_e$ dependence of the
normalized wave energy $\epsilon_p$ in red. We measured the average wave
energy inside the laminar wakefield
(e.g. $1350 \leq x/(c/\omega_{pe}) \leq 1850$ for $\sigma_e = 5$) in order
to evaluate the effect of the positive feedback
process on the wave emission efficiency and exclude the nonlinear effect of
the PDI. The dots and solid lines indicate
ion-electron and pair shock simulation results, respectively. We performed 1D
ion-electron shock simulations for comparison and the results are shown in
blue. The wave energy in 2D is systematically smaller
than that in 1D because of the inhomogeneity along the shock surface
\citep{Iwamoto2017,Iwamoto2018}.

The wave energy in ion-electron plasmas exceeds that in pair plasmas for both
1D and 2D except for the 2D run with $\sigma_e=0.1$. We attribute this
amplification over the pair plasmas to the positive feedback
process, which does not work in pair plasmas due to the absence of the
wakefield. If the electron kinetic energy achieves equipartition with the ion
during the positive feedback process, the incoming electron bulk Lorentz
factor becomes larger by a factor of $m_i/m_e$ than the initial. In other
words, the effective $\sigma_e$ decreases down to $(m_e/m_i)\sigma_e$,
whereas $\epsilon_p$ increases up to $(m_i/m_e)\epsilon_p$ during the
amplification phase. To confirm the hypothesis that the observed amplified
emission is due to
the positive feedback, we have also shown the emission efficiency measured in
pair plasmas rescaled by appropriate factors:
\begin{equation}
  \label{rescale}
  \epsilon_{p}^{\prime}(\sigma_e) =
  \frac{1}{2}\frac{m_i}{m_e}\epsilon_{p} \left( \frac{m_i}{m_e}\sigma_e\right)
\end{equation}
in left panel of Figure \ref{amp} with the dashed lines. Here the factor
$1/2$ is introduced because the wave amplitude in ion-electron plasmas was
obtained by averaging over the laminar-wakefield region. The wave emission
efficiency measured in pair plasmas $\epsilon_{p}$ can be converted to that
in ion-electron plasmas $\epsilon_{p}^{\prime}$ according to Eq.
\ref{rescale}. The $\sigma_e$ dependence of the wave energy in ion-electron
plasmas is in rough agreement with the positive feedback model associated
with the ion-electron coupling except for the 2D run with $\sigma_e = 0.1$.
Although the 2D run with $\sigma_e=1$ seems to be deviated from the rescaled
emission efficiency, the difference is at most a factor of three.
Furthermore, we investigated the mass ratio dependence
of the 1D wave energy. We fixed the upstream $\sigma_i=(m_e/m_i)\sigma_e$
and measured the wave energy $\epsilon_p$ in the same manner.
Our simulations in pair plasmas show that $\epsilon_p$, which is the
energy conversion rate from the free energy into the precursor wave energy,
is constant for a given $\sigma_e$. Since the free energy available for
the SMI increases with the mass ratio in ion-electron plasmas
due to the positive feedback process, the wave energy $\epsilon_p$ should be
a linear function of the mass ratio. As can be seen in right panel of
Figure \ref{amp}, $\epsilon_p$ is clearly proportional to $m_i/m_e$ for all
$\sigma_i$. This result also supports our model.

\added{Eq. \ref{rescale} indicates that the total magnetization
$\sigma_i$ dependence of the wave energy normalized by the total kinetic
energy $\delta B^2/4\pi \gamma_1N_1m_ic^2$ shows the same
tendency as that in pair plasmas. This is natural because the total kinetic
enegy is available for the SMI due to the ion-electron coupling. The
precursor wave emission is well-caracterized by the incoming total
kinetic energy as long as the positive feedback process works.}

In the case of the 2D shock at $\sigma_e=0.1$, we confirmed that the
wakefield is excited by the large-amplitude precursor waves and that the
wavelength of the laminar wakefield is consistent with the linear theory of
the PDI as with the case of relatively high magnetization. However, the
precursor wave amplitude is almost constant in the laminar-wakefield region
and the positive feedback does not work. One of the reason for this
discrepancy may be due to the inhomogeneity.
The ion-scale fluctuations along the shock surface are generated at
$\sigma_e=0.1$ \citep{Sironi2013} and the inhomogeneity at the shock front
is more prominent than
that for $\sigma_e >1$. Consequently, the wave emission via the SMI may
become less efficient at $\sigma_e=0.1$ \citep{Iwamoto2017,Iwamoto2018}.
For $\sigma_e >1$, the PDI rapidly grows
and induces the positive electric field in the immediate upstream at an early
stage of the shocks. Electrons enter the shock front during the acceleration
phase and the positive feedback process then begins. On the other hand, for
$\sigma_e = 0.1$, the precursor wave slowly decays into the wakefield
compared to that for $\sigma_e > 0.1$ because the linear growth rate of the
PDI is proportional to the amplitude of the pump wave \citep{Kruer1988}.
The wakefield is excited away from the shock front and the sinusoidal
electric field is generated even in an early phase. The net
acceleration of electrons is zero in the sinusoidal electric field and thus
the positive feedback does not work at $\sigma_e = 0.1$. If the positive
feedback process does not work, the wave emission in ion-electron plasmas
should be less efficient than that in pair plasmas because only electrons
emit the precursor waves via the SMI.

Although the wave emission efficiency deteriorates at $\sigma_e = 0.1$,
this may not be necessarily the case for a
sufficiently large mass ratio. We fixed the ion-to-electron mass ratio
$m_i/m_e=50$ throughout the 2D simulations. Since the ratio of the ion
gyroradius to the electron is equivalent to the mass ratio, the electron
gyromotion is nearly unperturbed in the ion-scale fluctuations for a
sufficiently large mass ratio and the wave emission via the SMI would
remain efficient at $\sigma_e = 0.1$. If such large mass ratio is used,
ion Weibel instability would dominate the shock front even for high
$\sigma_e$ and generate the ion-scale magnetic field
\citep{Sironi2013}. Electrons feel the Weibel-generated magnetic field and
may emit the stronger precursor waves \citep{Iwamoto2018}.
Therefore, we anticipate that the positive feedback process enhances the
precursor wave emission for a wide
range of $\sigma_e$ when the realistic mass ratio is used.

Since the incoming electrons are already thermalized before entering
into the shock, the wave amplitude may decrease in time or might be
completely shut off due to the suppression of the higher order harmonic
excitation \citep{Amato2006}. The weak precursor waves would not
excite the wakefields and the particle acceleration/heating may cease in a
later phase. After the quiescent stage of the precursor waves,
however, cold undisturbed electrons will
enter the shock once again and the whole positive feedback cycle will be
initiated. We thus speculate that this system may exhibit a cyclic behavior,
which may periodically induce the large-amplitude electromagnetic waves.

\section{Particle Acceleration}\label{sec:acc}

\begin{figure}[htb!]
 \plotone{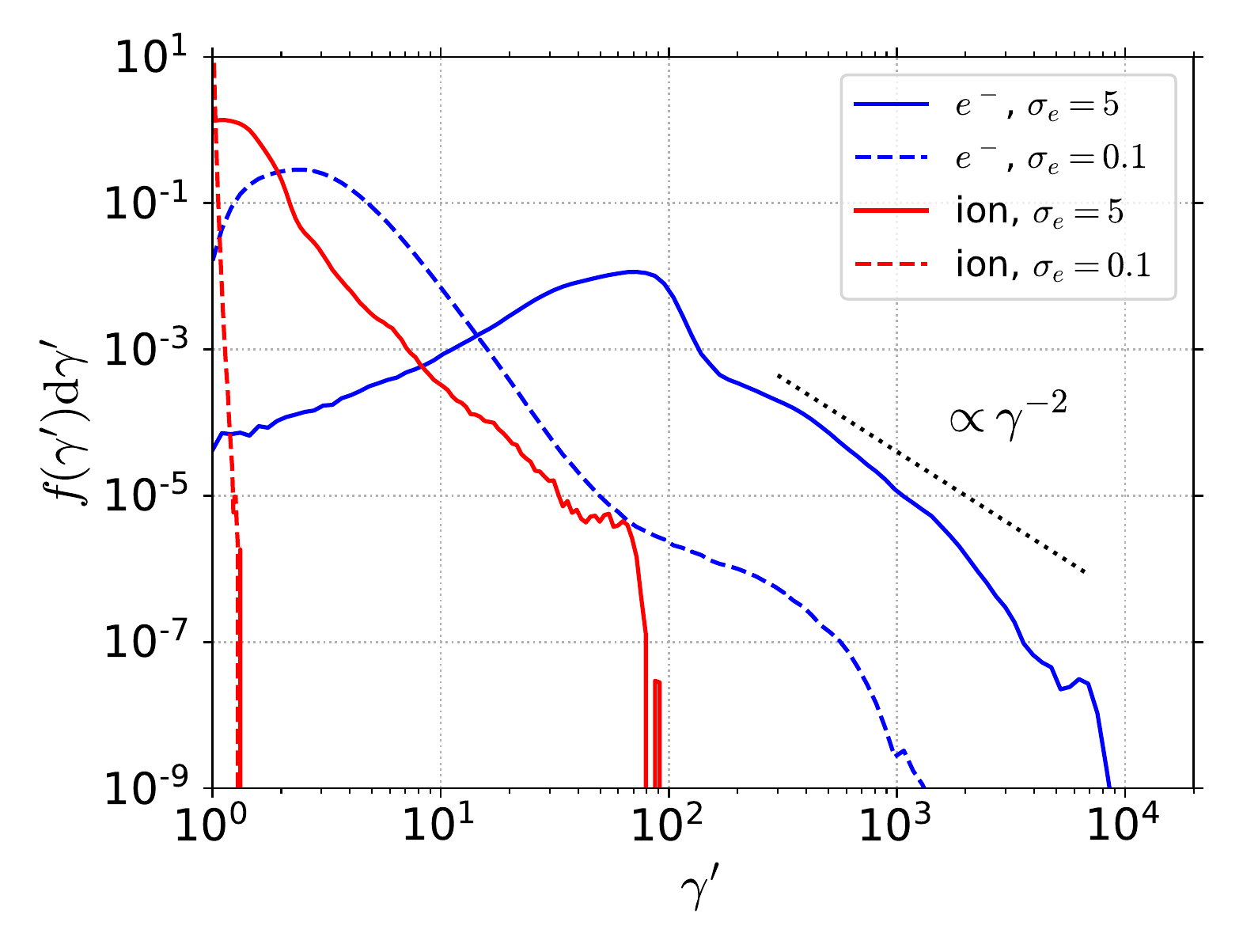}
 \caption{Near-upstream energy spectra of electrons (blue) and
 ions (red) for $\sigma_e = 5$ (solid lines) and $\sigma_e = 0.1$ (dashed
 lines) measured in the proper frame.}
 \label{dist}
\end{figure}

Figure \ref{dist} shows the energy spectra of electrons (blue)
and ions (red) in the turbulent-wakefield region, where the wakefield
collapses and the efficient particle acceleration occurs. The solid and
dashed lines indicate the spectra for $\sigma_e = 5$ and $\sigma_e = 0.1$,
respectively. The prime indicates physical quantities measured in the plasma
rest frame. Note that we use the proper frame for comparison with the
previous simulation results and laser experiments
\citep{Kuramitsu2011a,Kuramitsu2011b,Kuramitsu2008,
Kuramitsu2012,Liu2017,Liu2018,Liu2019}
where the spectra were measured in the plasma rest frame.
We determined the bulk velocity in the turbulent-wakefield region and then
the spectra are obtained by performing Lorentz transformation into the
plasma rest frame. A power-law distribution $\propto \gamma^{-2}$ are
also shown in black for reference. A clear nonthermal tail is observed
for electrons in the case of $\sigma_e = 5$, and the spectral
index is close to $2$. Surprisingly, nonthermal ions whose spectral
index is close to $2$ are also generated, which is a clear difference from
the earlier studies. At $\sigma_e = 0.1$, the wakefield amplitude is an order
of magnitude smaller than $\sigma_e=5$ and thus ions are almost unaffected by
such small wakefield. Electrons are accelerated, although less efficiently
compared to $\sigma_e=5$, and the spectrum has the nonthermal component.
The detailed acceleration mechanism will be presented in a future
publication.

As already discussed in Section \ref{sec:amp}, we expect the cyclic
excitation of the intense precursor waves in relativistic shocks.
We thus think that the particle acceleration is not transient and that
nonthermal particles may be periodically produced as well.

\section{Summary}\label{sec:sum}

In this work, we have found that the precursor
wave emission efficiency is dramatically enhanced for high $\sigma_e$ due to
the positive feedback process associated with the ion-electron
coupling. For low $\sigma_e$, the wave emission may be influenced by the
ion-scale fluctuations unless a sufficiently large mass ratio is applied.
The large-amplitude precursor wave initiates the nonlinear process of the PDI
and the wakefield is destroyed in the near-upstream region. Nonthermal
electrons and ions are generated inside the turbulent wakefield and
the energy spectrum exhibits a power-law distribution. This study shows that
the intense coherent wave emission and the particle acceleration can
operate in relativistic astrophysical objects.

\acknowledgments

 The authors would like to thank Shuichi Matsukiyo and Artem Bohdan for
 fruitful discussion.

 This work used the computational resources of the HPCI system provided
 by Information Technology Center, Nagoya University through the HPCI
 System Research Project (Project ID: hp180071,hp180180).

 This work is supported by Collaborative Research Project on Computer Science
 with High-Performance Computing in Nagoya University.

 Numerical computations were in part carried out on Cray XC50 at
 Center for Computational Astrophysics, National Astronomical
 Observatory of Japan.

 This work was supported in part by JSPS KAKENHI grant No. 17H02877.

 The work of J. N., A. L., and O. K. has been supported by Narodowe Centrum
 Nauki through research project DEC-2013/10/E/ST9/00662.


\begin{thebibliography}{}
 \expandafter\ifx\csname natexlab\endcsname\relax\def\natexlab#1{#1}\fi
 \providecommand{\url}[1]{\href{#1}{#1}}
 \providecommand{\dodoi}[1]{doi:~\href{http://doi.org/#1}{\nolinkurl{#1}}}
 \providecommand{\doeprint}[1]{\href{http://ascl.net/#1}{\nolinkurl{http://ascl.net/#1}}}
 \providecommand{\doarXiv}[1]{\href{https://arxiv.org/abs/#1}{\nolinkurl{https://arxiv.org/abs/#1}}}

 \bibitem[{Aab {et~al.}(2015)Aab, Abreu, Aglietta, Ahn, Samarai, Albuquerque,
   Allekotte, Allen, Allison, Almela, {et~al.}}]{Aab2015}
 Aab, A., Abreu, P., Aglietta, M., {et~al.} 2015, \apj, 804, 15,
   \dodoi{10.1088/0004-637X/804/1/15}

 \bibitem[{Aab {et~al.}(2017)Aab, Abreu, Aglietta, {Al Samarai}, Albuquerque,
   Allekotte, Almela, {Alvarez Castillo}, Alvarez-Mu{\~{n}}iz, Anastasi,
   {et~al.}}]{Aab2017}
 ---. 2017, Sci, 357, 1266, \dodoi{10.1126/science.aan4338}

 \bibitem[{Aab {et~al.}(2018)Aab, Abreu, Aglietta, Albuquerque, Allekotte,
   Almela, Castillo, Alvarez-Mu{\~{n}}iz, Anastasi, Anchordoqui,
   {et~al.}}]{Aab2018}
 ---. 2018, \apjl, 853, L29, \dodoi{10.3847/2041-8213/aaa66d}

 \bibitem[{Aartsen {et~al.}(2018)Aartsen, Ackermann, Adams, Aguilar, Ahlers,
   Ahrens, Samarai, Altmann, Andeen, Anderson, {et~al.}}]{Aartsen2018}
 Aartsen, M.~G., Ackermann, M., Adams, J., {et~al.} 2018, Sci, 361, 147,
   \dodoi{10.1126/science.aat2890}

 \bibitem[{Abbasi {et~al.}(2014)Abbasi, Abe, Abu-Zayyad, Allen, Anderson, Azuma,
   Barcikowski, Belz, Bergman, Blake, {et~al.}}]{Abbasi2014}
 Abbasi, R.~U., Abe, M., Abu-Zayyad, T., {et~al.} 2014, \apjl, 790, L21,
   \dodoi{10.1088/2041-8205/790/2/L21}

 \bibitem[{Amato \& Arons(2006)}]{Amato2006}
 Amato, E., \& Arons, J. 2006, \apj, 653, 325, \dodoi{10.1086/508050}

 \bibitem[{Arons(2003)}]{Arons2003}
 Arons, J. 2003, \apj, 589, 871, \dodoi{10.1086/374776}

 \bibitem[{Biermann \& Strittmatter(1987)}]{Biermann1987}
 Biermann, P.~L., \& Strittmatter, P.~A. 1987, apj, 322, 643,
   \dodoi{10.1086/165759}

 \bibitem[{Chen {et~al.}(2002)Chen, Tajima, \& Takahashi}]{Chen2002}
 Chen, P., Tajima, T., \& Takahashi, Y. 2002, \prl, 89, 161101,
   \dodoi{10.1103/PhysRevLett.89.161101}

 \bibitem[{Fried(1959)}]{Fried1959}
 Fried, B.~D. 1959, PhFl, 2, 337, \dodoi{10.1063/1.1705933}

 \bibitem[{Gallant {et~al.}(1992)Gallant, Hoshino, Langdon, Arons, \&
   Max}]{Gallant1992}
 Gallant, Y.~A., Hoshino, M., Langdon, A.~B., Arons, J., \& Max, C.~E. 1992,
   \apj, 391, 73, \dodoi{10.1086/171326}

 \bibitem[{Hillas(1984)}]{Hillas1984}
 Hillas, A.~M. 1984, \araa, 22, 425, \dodoi{10.1146/annurev.aa.22.090184.002233}

 \bibitem[{Hoshino(2008)}]{Hoshino2008}
 Hoshino, M. 2008, \apj, 672, 940, \dodoi{10.1086/523665}

 \bibitem[{Hoshino \& Arons(1991)}]{Hoshino1991}
 Hoshino, M., \& Arons, J. 1991, PhFlB, 3, 818, \dodoi{10.1063/1.859877}

 \bibitem[{Hoshino {et~al.}(1992)Hoshino, Arons, Gallant, \&
   Langdon}]{Hoshino1992}
 Hoshino, M., Arons, J., Gallant, Y.~A., \& Langdon, A.~B. 1992, \apj, 390, 454,
   \dodoi{10.1086/171296}

 \bibitem[{Ikeya \& Matsumoto(2015)}]{Ikeya2015}
 Ikeya, N., \& Matsumoto, Y. 2015, \pasj, 67, 64, \dodoi{10.1093/pasj/psv052}

 \bibitem[{Iwamoto {et~al.}(2017)Iwamoto, Amano, Hoshino, \&
   Matsumoto}]{Iwamoto2017}
 Iwamoto, M., Amano, T., Hoshino, M., \& Matsumoto, Y. 2017, \apj, 840, 52,
   \dodoi{10.3847/1538-4357/aa6d6f}

 \bibitem[{Iwamoto {et~al.}(2018)Iwamoto, Amano, Hoshino, \&
   Matsumoto}]{Iwamoto2018}
 ---. 2018, \apj, 858, 93, \dodoi{10.3847/1538-4357/aaba7a}

 \bibitem[{Kruer(1988)}]{Kruer1988}
 Kruer, W.~L. 1988, The Physics of Laser Plasma Interactions, ed. D.~Pines
   (Boston: Addison-Wesley)

 \bibitem[{Kuramitsu {et~al.}(2012)Kuramitsu, Sakawa, Hoshino, Chen, \&
   Takabe}]{Kuramitsu2012}
 Kuramitsu, Y., Sakawa, Y., Hoshino, M., Chen, S.~H., \& Takabe, H. 2012, HEDP,
   8, 266, \dodoi{10.1016/j.hedp.2012.03.016}

 \bibitem[{Kuramitsu {et~al.}(2008)Kuramitsu, Sakawa, Kato, Takabe, \&
   Hoshino}]{Kuramitsu2008}
 Kuramitsu, Y., Sakawa, Y., Kato, T., Takabe, H., \& Hoshino, M. 2008, \apj,
   682, 113, \dodoi{10.1086/591247}

 \bibitem[{Kuramitsu {et~al.}(2011{\natexlab{a}})Kuramitsu, Nakanii, Kondo,
   Sakawa, Mori, Miura, Tsuji, Kimura, Fukumochi, Kashihara,
   {et~al.}}]{Kuramitsu2011a}
 Kuramitsu, Y., Nakanii, N., Kondo, K., {et~al.} 2011{\natexlab{a}}, PhPl, 18,
   010701, \dodoi{10.1063/1.3528434}

 \bibitem[{Kuramitsu {et~al.}(2011{\natexlab{b}})Kuramitsu, Nakanii, Kondo,
   Sakawa, Mori, Miura, Tsuji, Kimura, Fukumochi, Kashihara,
   {et~al.}}]{Kuramitsu2011b}
 ---. 2011{\natexlab{b}}, \pre, 83, 026401, \dodoi{10.1103/PhysRevE.83.026401}

 \bibitem[{Langdon {et~al.}(1988)Langdon, Arons, \& Max}]{Langdon1988}
 Langdon, A.~B., Arons, J., \& Max, C.~E. 1988, \prl, 61, 779,
   \dodoi{10.1103/PhysRevLett.61.779}

 \bibitem[{Liu {et~al.}(2019)Liu, Isayama, Chen, \& Kuramitsu}]{Liu2019}
 Liu, Y.~L., Isayama, S., Chen, S.~H., \& Kuramitsu, Y. 2019, HEDP, 31, 64,
   \dodoi{10.1016/j.hedp.2019.03.004}

 \bibitem[{Liu {et~al.}(2018)Liu, Kuramitsu, Isayama, \& Chen}]{Liu2018}
 Liu, Y.~L., Kuramitsu, Y., Isayama, S., \& Chen, S.~H. 2018, PhPl, 25, 013110,
   \dodoi{10.1063/1.5006325}

 \bibitem[{Liu {et~al.}(2017)Liu, Kuramitsu, Moritaka, \& Chen}]{Liu2017}
 Liu, Y.~L., Kuramitsu, Y., Moritaka, T., \& Chen, S.~H. 2017, HEDP, 22, 46,
   \dodoi{10.1016/j.hedp.2017.02.006}

 \bibitem[{Lyubarsky(2006)}]{Lyubarsky2006}
 Lyubarsky, Y. 2006, \apj, 652, 1297, \dodoi{10.1086/508606}

 \bibitem[{Lyubarsky(2014)}]{Lyubarsky2014}
 ---. 2014, \mnras, 442, L9, \dodoi{10.1093/mnrasl/slu046}

 \bibitem[{Lyubarsky(2018)}]{Lyubarsky2018}
 ---. 2018, \mnras, 474, 1135, \dodoi{10.1093/mnras/stx2832}

 \bibitem[{Marscher(2006)}]{Marscher2006}
 Marscher, A.~P. 2006, in AIP Conf. Proc., Vol. 856, RELATIVISTIC JETS: The
   Common Physics of AGN, Microquasars, and Gamma-Ray Bursts, ed. P.~A. Hughes
   \& J.~N. Bregman (Ann Arbor, Michigan: AIP), 1--22

 \bibitem[{Matsumoto {et~al.}(2013)Matsumoto, Amano, \& Hoshino}]{Matsumoto2013}
 Matsumoto, Y., Amano, T., \& Hoshino, M. 2013, \prl, 111, 215003,
   \dodoi{10.1103/PhysRevLett.111.215003}

 \bibitem[{Matsumoto {et~al.}(2015)Matsumoto, Amano, Kato, \&
   Hoshino}]{Matsumoto2015}
 Matsumoto, Y., Amano, T., Kato, T.~N., \& Hoshino, M. 2015, Sci, 347, 974,
   \dodoi{10.1126/science.1260168}

 \bibitem[{Metzger {et~al.}(2019)Metzger, Margalit, \& Sironi}]{Metzger2019}
 Metzger, B.~D., Margalit, B., \& Sironi, L. 2019, \mnras, 485, 4091,
   \dodoi{10.1093/mnras/stz700}

 \bibitem[{Milgrom \& Usov(1995)}]{Milgrom1995}
 Milgrom, M., \& Usov, V. 1995, \apjl, 449, L37, \dodoi{10.1086/309633}

 \bibitem[{Mima \& Nishikawa(1984)}]{Mima1984}
 Mima, K., \& Nishikawa, K. 1984, Basic Plasma Physics, ed. A.~A. Galeev \&
   R.~N. Sudan (Amsterdam: North-Holland Publishing Company)

 \bibitem[{Murase {et~al.}(2009)Murase, M{\'{e}}sz{\'{a}}ros, \&
   Zhang}]{Murase2009}
 Murase, K., M{\'{e}}sz{\'{a}}ros, P., \& Zhang, B. 2009, \prd, 79, 103001,
   \dodoi{10.1103/PhysRevD.79.103001}

 \bibitem[{Piran(2005)}]{Piran2005}
 Piran, T. 2005, \rmp, 76, 1143, \dodoi{10.1103/RevModPhys.76.1143}

 \bibitem[{Plotnikov {et~al.}(2018)Plotnikov, Grassi, \& Grech}]{Plotnikov2018}
 Plotnikov, I., Grassi, A., \& Grech, M. 2018, \mnras, 477, 5238,
   \dodoi{10.1093/mnras/sty979}

 \bibitem[{Plotnikov \& Sironi(2019)}]{Plotnikov2019}
 Plotnikov, I., \& Sironi, L. 2019, \mnras, 485, 3816,
   \dodoi{10.1093/mnras/stz640}

 \bibitem[{Sironi {et~al.}(2013)Sironi, Spitkovsky, \& Arons}]{Sironi2013}
 Sironi, L., Spitkovsky, A., \& Arons, J. 2013, \apj, 771, 54,
   \dodoi{10.1088/0004-637X/771/1/54}

 \bibitem[{Tajima \& Dawson(1979)}]{Tajima1979}
 Tajima, T., \& Dawson, J.~M. 1979, \prl, 43, 267,
   \dodoi{10.1103/PhysRevLett.43.267}

 \bibitem[{Vietri(1995)}]{Vietri1995}
 Vietri, M. 1995, \apj, 453, 883, \dodoi{10.1086/176448}

 \bibitem[{Waxman(1995)}]{Waxman1995}
 Waxman, E. 1995, \apjl, 452, L1, \dodoi{10.1086/309715}

 \bibitem[{Weibel(1959)}]{Weibel1959}
 Weibel, Erich, S. 1959, \prl, 2, 83, \dodoi{10.1103/PhysRevLett.2.83}

 \end{thebibliography}

\end{document}